\documentclass[sigconf,nonacm]{acmart}
\AtBeginDocument{%
  }

\setcopyright{none}
\usepackage{enumitem} 
\usepackage{graphicx} 

\usepackage{booktabs}
\usepackage{xcolor}
\usepackage{colortbl}
\usepackage{pifont}

\usepackage{tikz}
\newcommand*\emptycirc[1][0.8ex]{\tikz\draw (0,0) circle (#1);}
\newcommand*\halfcirc[1][0.8ex]{%
 \begin{tikzpicture}
 \draw[fill] (0,0)-- (90:#1) arc (90:270:#1) -- cycle ;
 \draw (0,0) circle (#1);
 \end{tikzpicture}}
\newcommand*\fullcirc[1][0.8ex]{\tikz\fill (0,0) circle (#1);}

\usepackage{multirow}  
\usepackage{threeparttable}
\usepackage{booktabs}

\usepackage{booktabs}
\usepackage{pifont}

\newcommand{\yesmark}{\ding{51}}
\newcommand{\nomark}{\ding{55}}

\usepackage[table]{xcolor}  
\usepackage{xspace}

\begin{document}

\title{Tabular Foundation Models for Multi-View Information Cascade Popularity Prediction}



\author{Wenting Zhu}
\affiliation{%
  \institution{Beijing University of Posts and Telecommunications}
  \city{Beijing}
  \country{China}
  }
\email{zwt@bupt.edu.cn}

\author{Chenghua Gong}
\affiliation{%
  \institution{University of Science and Technology of China}
  \city{Hefei}
  \country{China}
  }
\email{gongchenghua@mail.ustc.edu.cn}

\author{Sanchuan Guo}
\affiliation{%
  \institution{Beijing University of Posts and Telecommunications}
  \city{Beijing}
  \country{China}
  }
\email{guosc@bupt.edu.cn}

\author{Chaozhuo Li}
\affiliation{%
  \institution{Beijing University of Posts and Telecommunications}
  \city{Beijing}
  \country{China}
  }
\email{lichaozhuo@bupt.edu.cn}

\author{Yueyue Zhang}
\affiliation{%
  \institution{Beijing University of Posts and Telecommunications}
  \city{Beijing}
  \country{China}
  }
\email{zhangyueyue@bupt.edu.cn}

\author{Xi Zhang}
\affiliation{%
  \institution{Beijing University of Posts and Telecommunications}
  \city{Beijing}
  \country{China}
  }
\email{zhangx@bupt.edu.cn}

\renewcommand{\shortauthors}{Zhu et al.}

\begin{abstract}

Predicting the future popularity of information cascades is essential for understanding information diffusion on social media. Despite recent advances, existing methods face two key limitations: they focus primarily on the cascade view while overlooking other information views that drive user engagement, such as textual semantics, visual content, and tabular attributes; and they fail to capture high-order cross-view interactions. To address these issues, we propose \textbf{TFM4POP}, the first framework to introduce tabular foundation models (TFMs) into popularity prediction, leveraging their pre-trained tabular priors to unify the modeling of multiple heterogeneous information views. Specifically, TFM4POP adopts a dual-branch design: the static branch employs a TFM as the feature-encoding backbone that jointly reasons over all static views through in-context learning to produce the static cascade representation, while the dynamic branch captures the continuous-time cascade dynamics with a dedicated Neural-ODE-based encoder. The two representations are then fused via cross-attention for the final prediction. Furthermore, to adapt the TFM to real cascade distributions, we apply parameter-efficient IA3 fine-tuning, achieving performance competitive with or better than full fine-tuning while updating substantially fewer parameters. In addition, we construct a comprehensive multi-view cascade benchmark that covers all four information views. Extensive experiments show that TFM4POP consistently outperforms state-of-the-art baselines across multiple datasets and observation settings.

\end{abstract}







\maketitle

\section{Introduction}
\label{sec:intro}
\begin{figure}[t]
\centering
\includegraphics[trim=3 2 8 1, clip,  width=\columnwidth]{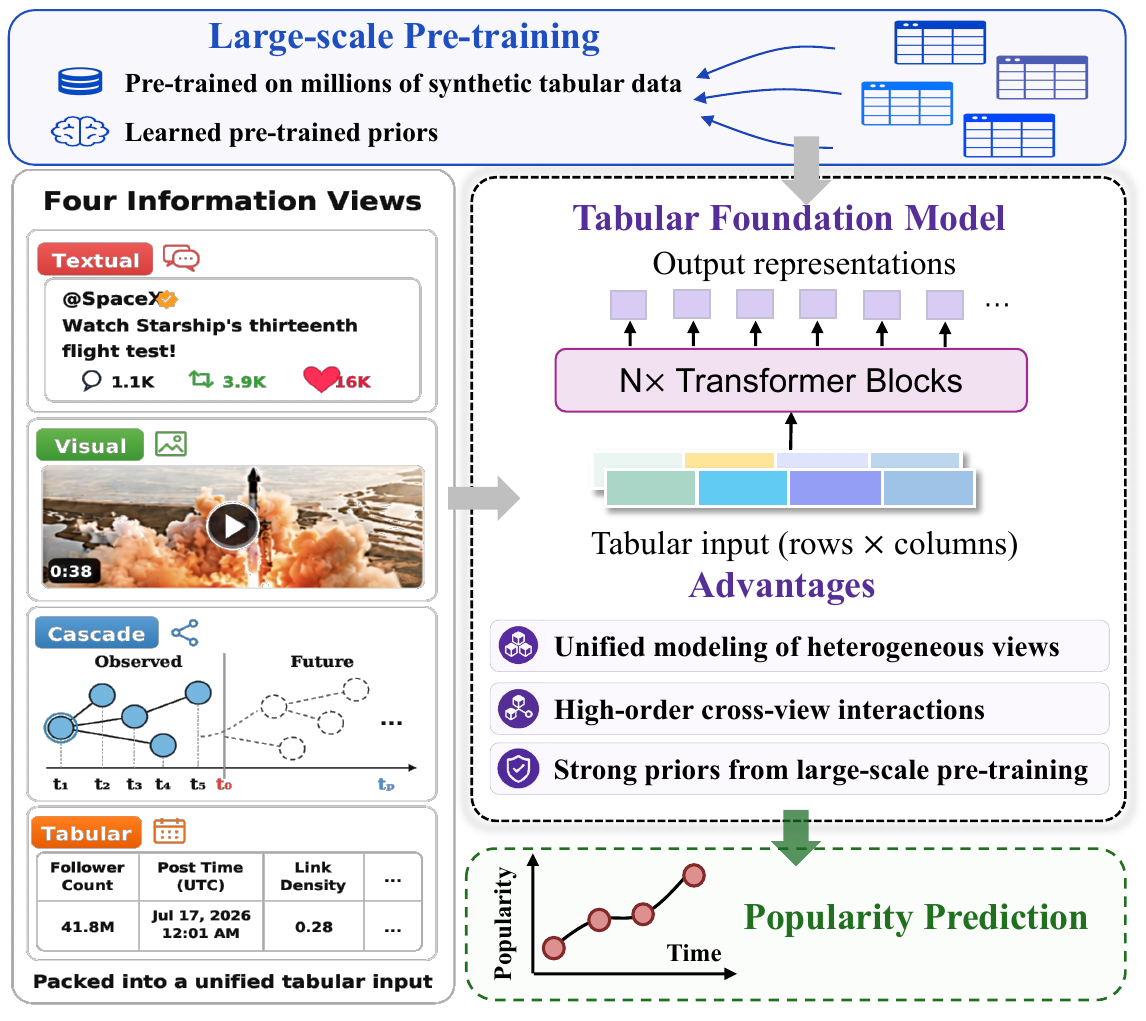}
\vspace{-5mm}  
\caption{TFM4POP packs four information views into a unified tabular input, letting a pre-trained TFM jointly reason over all of them.}
\label{fig:introduction}
\vspace{-5mm}  
\end{figure}

\textbf{Background.} Online social platforms have brought unprecedented convenience for the production and dissemination of information, while also amplifying the rapid spread of misinformation \cite{budak2011limiting, sun2023fighting}, hate speech \cite{masud2021hate, goel2023hatemongers}, and other harmful narratives \cite{mladenovic2021cyber}. In this context, understanding and modeling how information propagates on social platforms has emerged as a hot research topic, with wide applications including misinformation control \cite{liu2018early, sharma2021network}, trending topic detection \cite{miao2016cost}, and online marketing \cite{aggrawal2017brand}.  
A key research task in this field is information cascade popularity prediction, which aims to forecast the future popularity of a cascade from its early-stage diffusion patterns \cite{cheng2024information}. 
As shown in Figure~\ref{fig:introduction}, post popularity is jointly shaped by heterogeneous signals spanning four information views: textual content, visual content, the cascade view, and tabular attributes. The last of these comprises readily available metadata, such as user profiles and posting time, together with hand-crafted statistical features whose predictive value is well established in early feature-engineering studies~\cite{cheng2014can, shulman2016predictability}.

Existing methods fall into the following four categories \cite{cheng2024information, bao2024popularity}. (1) Feature-based methods \cite{tsur2012s, romero2013interplay, cheng2014can, shulman2016predictability} extract various hand-crafted features related to users, content, cascade structure, and temporal dynamics. (2) Statistics-based methods characterize cascade growth with Poisson~\cite{shen2014modeling} or Hawkes~\cite{cao2017deephawkes, rizoiu2017expecting} point processes. (3) Deep learning-based methods focus on learning expressive cascade representations using RNNs \cite{li2017deepcas, chen2019information}, GNNs \cite{cao2020popularity, bao2024popularity}, Transformers \cite{chen2022and, wang2025casformer}, and Neural ODEs \cite{cheng2024information}. Owing to their superior performance, these approaches have become the dominant paradigm.
(4) LLM-based methods have recently emerged, which leverage the strong reasoning and semantic understanding capabilities of large language models (LLMs) to generate auxiliary insights \cite{xu2025forecasting,kayal2025large,shangmake} or directly predict popularity \cite{zheng2025autocas}. 
Despite these significant advances, most existing methods still suffer from two major limitations.

\textbf{(L1) Incomplete view coverage and shallow fusion.}
As summarized in Table~\ref{tab:comparison}, no existing method effectively models all four views simultaneously; most focus primarily on the cascade view.
Worse still, even when multiple views are considered, existing models typically encode each view independently and fuse them through shallow strategies such as feature concatenation or gating, which fail to capture high-order cross-view interactions---for example, how the appeal of the content is amplified or muted by who posts it and when.
Modeling such interactions is essential not only for capturing \textit{how} information spreads, but also for understanding \textit{why} it resonates with the audience.
\textbf{(L2) Inability to exploit large-scale pre-trained priors.} 
In contrast to the pretrain-then-adapt paradigm that has proven transformative in language and vision, existing approaches are trained from scratch on individual, moderately sized cascade datasets, and thus cannot benefit from the knowledge acquired through large-scale pre-training. This raises a natural question: \textit{can cascade popularity prediction benefit from the same paradigm?} Recent LLM-based methods have made preliminary attempts in this direction. Yet LLMs derive their priors predominantly from text corpora, and struggle with the structured tabular and graph data in which most of the diffusion signal resides. We argue that this task instead calls for a backbone pre-trained on heterogeneous structured data, capable of jointly encoding the multiple information views above within a single, unified framework.

Tabular foundation models (TFMs) provide such a backbone. TFMs like TabPFN~\cite{hollmann2022tabpfn, hollmann2025accurate} and LimiX~\cite{zhang2025limix} treat supervised learning on tables as amortized Bayesian inference: pre-trained on millions of synthetic datasets drawn from a predefined prior, they approximate the Bayesian posterior predictive distribution and predict on unseen datasets via in-context learning in one forward pass, without parameter updates. They alternate row-wise inter-feature attention with column-wise inter-sample attention, modeling dependencies among features within a sample and among samples for a given feature, and natively accommodate heterogeneous feature types. This suggests one move that addresses both limitations at once: if the non-tabular views are transformed into compact feature columns, all heterogeneous signals reside in one shared tabular space, where row-wise attention performs high-order cross-view fusion (L1), while the pre-trained prior, together with column-wise attention relating each cascade to similar historical ones, supplies knowledge that the cascade corpus itself lacks (L2).

\begin{table}[t]
\centering
\caption{Comparison of our proposed TFM4POP with existing popularity prediction approaches. \fullcirc\ indicates the view is fully modeled, \halfcirc\ partially modeled, and \emptycirc\ not modeled.}
\label{tab:comparison}
\setlength{\tabcolsep}{4pt}          
\renewcommand{\arraystretch}{0.75}   
\begin{tabular}{lcccc}
\toprule
\textbf{Model} & \textbf{Textual} & \textbf{Visual} & \textbf{Cascade} & \textbf{Tabular} \\
\midrule
UHAN~\cite{zhang2018user}           & \fullcirc  & \fullcirc  & \emptycirc & \emptycirc \\
CBAN~\cite{cheung2022crossmodal}    & \fullcirc  & \fullcirc  & \emptycirc & \emptycirc \\
DeepCas~\cite{li2017deepcas}        & \emptycirc & \emptycirc & \halfcirc  & \emptycirc \\
DeepHawkes~\cite{cao2017deephawkes} & \emptycirc & \emptycirc & \halfcirc  & \emptycirc \\
CasFlow~\cite{xu2021casflow}        & \emptycirc & \emptycirc & \fullcirc  & \emptycirc \\
CTCP~\cite{lu2023continuous}        & \emptycirc & \emptycirc & \fullcirc  & \emptycirc \\
CasDo~\cite{cheng2024information}   & \emptycirc & \emptycirc & \fullcirc  & \emptycirc \\
CasFT~\cite{jing2025casft}          & \emptycirc & \emptycirc & \fullcirc  & \emptycirc \\
BuzzProphet~\cite{jing2026modeling} & \fullcirc  & \emptycirc & \emptycirc & \emptycirc \\
AutoCas~\cite{zheng2025autocas}       & \emptycirc  & \emptycirc  & \fullcirc  & \emptycirc  \\
MMCas~\cite{jing2026modeling}       & \fullcirc  & \fullcirc  & \fullcirc  & \halfcirc  \\
\midrule
\textbf{TFM4POP (ours)}             & \fullcirc  & \fullcirc  & \fullcirc  & \fullcirc  \\
\bottomrule
\end{tabular}
\vspace{-3mm}
\end{table}

\textbf{Challenges.} However, designing an effective TFM-based framework for popularity prediction is non-trivial and poses three challenges.
\textbf{(C1) Modality gap.}
TFMs are pre-trained exclusively on synthetic tabular data, and their principled extension to text, images, and graph structures remains underexplored. Naively concatenating the raw high-dimensional feature vectors of non-tabular views with the tabular attributes as the TFM input may cause attention imbalance: views occupying more feature columns dominate the attention budget and suppress the tabular signals.
\textbf{(C2) Integration of cascade dynamics.}
Cascade diffusion is a continuous-time event sequence in which reposts arrive at irregular intervals, and thus cannot be faithfully summarized as a static table row. How to effectively capture such temporal information and fuse it with the static views remains a challenge.
\textbf{(C3) Adaptation dilemma.}
Real cascade distributions inevitably deviate from the synthetic pre-training distribution, yet naive fine-tuning risks catastrophically forgetting the pre-trained prior. It remains unclear whether fine-tuning is beneficial and which strategies (parameter-efficient fine-tuning (PEFT) or full fine-tuning) are effective.

To address these challenges, we propose \textbf{TFM4POP}, a novel framework that repurposes a TFM as a unified multi-view fusion backbone for popularity prediction.
For \textbf{C1}, we design a static branch that first encodes each non-tabular view into a few compact feature columns and then assembles them with the raw tabular attributes into a unified tabular input, enabling the TFM to jointly reason over all static views with its pre-trained prior while avoiding attention imbalance across views.
For \textbf{C2}, we devise a dynamic branch with a Neural-ODE-based encoder to model the irregular, continuous-time repost sequences. The resulting dynamic representations are then fused with the static representations via cross-attention for prediction. 
For \textbf{C3}, we compare different adaptation strategies and adopt parameter-efficient IA3 fine-tuning, which preserves the pre-trained prior and achieves performance competitive with full fine-tuning while updating substantially fewer trainable parameters.
Furthermore, given the scarcity of multi-view cascade benchmarks, we construct \textbf{EventCas}, a new event-centric cascade dataset that covers all four information views. As shown in Table~\ref{tab:dataset_comparison}, EventCas provides more comprehensive information coverage than existing benchmarks while remaining substantial in scale, enabling a rigorous validation of our motivation and design choices.

\textbf{Contributions.} The main contributions of our paper can be summarized as follows:
\begin{itemize}[leftmargin=*]

\item We propose TFM4POP, to our knowledge the first framework to introduce TFMs into popularity prediction. By transforming the heterogeneous information views into a unified tabular input, TFM4POP enables a TFM to jointly reason over all views with its large-scale pre-trained priors, achieving high-order cross-view fusion and more accurate prediction.






\item To adapt the TFM to real cascade distributions while preserving its pretrained prior, we compare a range of adaptation strategies (freezing, PEFT, or full fine-tuning) and find that PEFT, especially IA3, achieves performance on par with, or better than, full fine-tuning while updating substantially fewer trainable parameters. These findings offer valuable insights for extending TFMs beyond standard tabular data.


\item We construct a new event-centric multi-view cascade benchmark that offers more comprehensive information coverage than existing ones (Table~\ref{tab:dataset_comparison}), enabling more realistic and context-aware popularity prediction. The benchmark will be publicly released to facilitate future research on information cascades.


\item Extensive experiments on two benchmark datasets show that TFM4POP consistently outperforms state-of-the-art baselines, and ablation studies validate each design choice.

\end{itemize}

\section{Related Work}
\noindent \textbf{Information Cascade Popularity Prediction.}
Existing methods fall into four categories. Feature-based methods \cite{tsur2012s, romero2013interplay, cheng2014can, shulman2016predictability} rely on hand-crafted features, whose performance hinges on feature quality and generalizes poorly. Statistics-based methods characterize diffusion with Poisson \cite{shen2014modeling} or Hawkes \cite{cao2017deephawkes, rizoiu2017expecting} processes, offering interpretability but limited accuracy. Deep learning-based methods \cite{chen2019information, cao2020popularity, xu2021casflow, lu2023continuous, ji2023community, bao2024popularity, cheng2024information, wang2025casformer} combine graph and sequence models to learn expressive cascade representations and have become the dominant paradigm. More recently, LLM-based methods \cite{xu2025forecasting, kayal2025large, shangmake, zheng2025autocas} leverage LLMs to generate auxiliary insights or directly predict popularity. Despite these advances, most methods focus primarily on the cascade view while under-exploiting textual, visual, and tabular signals; even when multiple views are available, they are fused only shallowly (e.g., by concatenation or gating). Moreover, these methods are trained from scratch on moderately sized cascade datasets, and thus cannot leverage large-scale structured priors.

\noindent \textbf{Tabular Foundation Models.}
TFMs extend the foundation-model paradigm to tabular data: pre-trained on massive synthetic datasets, models such as TabPFN \cite{hollmann2022tabpfn, hollmann2025accurate} and LimiX \cite{zhang2025limix} generalize to unseen datasets via in-context learning (Section~\ref{sec:pre_tfm}). TFMs have recently attracted growing interest and been successfully applied to time series forecasting \cite{hoo2024tabular, hoo2025tables}, graph anomaly detection \cite{liu2026tabular}, and medical diagnosis \cite{rawat2025fast,pinero2026taco}. However, applying TFMs to popularity prediction is non-trivial: they cannot directly ingest raw text, images, or continuous-time cascade dynamics, and their synthetic pre-training prior may deviate from real-world cascade distributions. To this end, we propose TFM4POP, which, to our knowledge, is the first to repurpose a TFM as a unified multi-view fusion backbone for popularity prediction.


\section{Preliminaries}

\subsection{Problem Formulation}

\noindent \textbf{Definition 1. (Information Cascade)} : 
Given a user set $\mathcal{U}$, a cascade (or cascade sequence) $C_i=\{(u_j,t_j)\}_{j=1}^{|C_i|}$ records the diffusion process of post $i$ in chronological order. The root tuple $(u_1,t_1)$ denotes the original publication, and each subsequent tuple $(u_j,t_j)$ indicates that user $u_j$ reposts the post at timestamp $t_j$.

\noindent \textbf{Definition 2. (Cascade Graph)} : 
Given a cascade $C_i(t_o)$ observed up to time $t_o$, its propagation structure is represented as a directed graph $G_i(t_o)=(V_i(t_o),E_i(t_o))$. Here, $V_i(t_o)$ denotes the set of users participating in the diffusion of post $i$ by time $t_o$, and each edge $e_{j,k}=(u_j,u_k)\in E_i(t_o)$ indicates that user $u_k$ reposts the post from user $u_j$ at timestamp $t_k\leq t_o$. Fig.~\ref{fig:introduction} illustrates how a cascade graph evolves over time.

\noindent \textbf{Definition 3. (Global Graph)} :
Following prior work~\cite{jing2025casft, jing2026modeling}, we construct a directed global graph $\mathcal{G} = (\mathcal{V}, \mathcal{E})$ from all cascades observed by time $t_o$. Each edge in $\mathcal{E}$ represents a user-to-user diffusion relationship observed across cascades.

\noindent \textbf{Definition 4. (Static Attributes)} :
Each cascade $C_i$ is further associated with a set of static attributes $\mathcal{X}_i=\{\mathbf{x}_i^{\mathrm{tab}},\mathcal{T}_i,\mathcal{I}_i\}$. The tabular attributes $\mathbf{x}_i^{\mathrm{tab}}$ comprise metadata associated with post $i$, such as the profiles of the poster and early participants and the posting time, together with hand-crafted statistical features extracted from the observed cascade. $\mathcal{T}_i$ and $\mathcal{I}_i$ denote the textual and visual content of the post, respectively. Together with the cascade view captured by $C_i$ and $G_i$, these attributes constitute four complementary information views of post $i$. Visual content may be unavailable for posts without an attached image or video.

\noindent \textbf{Definition 5. (Popularity Prediction)} : 
Given a cascade $C_i(t_o)$ observed up to the observation time $t_o$, together with its cascade graph $G_i(t_o)$, the global graph $\mathcal{G}$, and the static attributes $\mathcal{X}_i$, our goal is to predict the incremental popularity of post $i$ at a future prediction time $t_p\gg t_o$. Let $|C_i(t)|$ denote the number of diffusion events observed in cascade $C_i$ by time $t$. The incremental popularity is then defined as $\Delta P_i=|C_i(t_p)|-|C_i(t_o)|$. We therefore formulate cascade popularity prediction as a regression task that learns a function $f:(C_i(t_o),G_i(t_o),\mathcal{G},\mathcal{X}_i)\rightarrow\Delta P_i$.

\begin{figure}[t]
\centering
\includegraphics[trim=0 0 0 0, clip, width=\columnwidth]{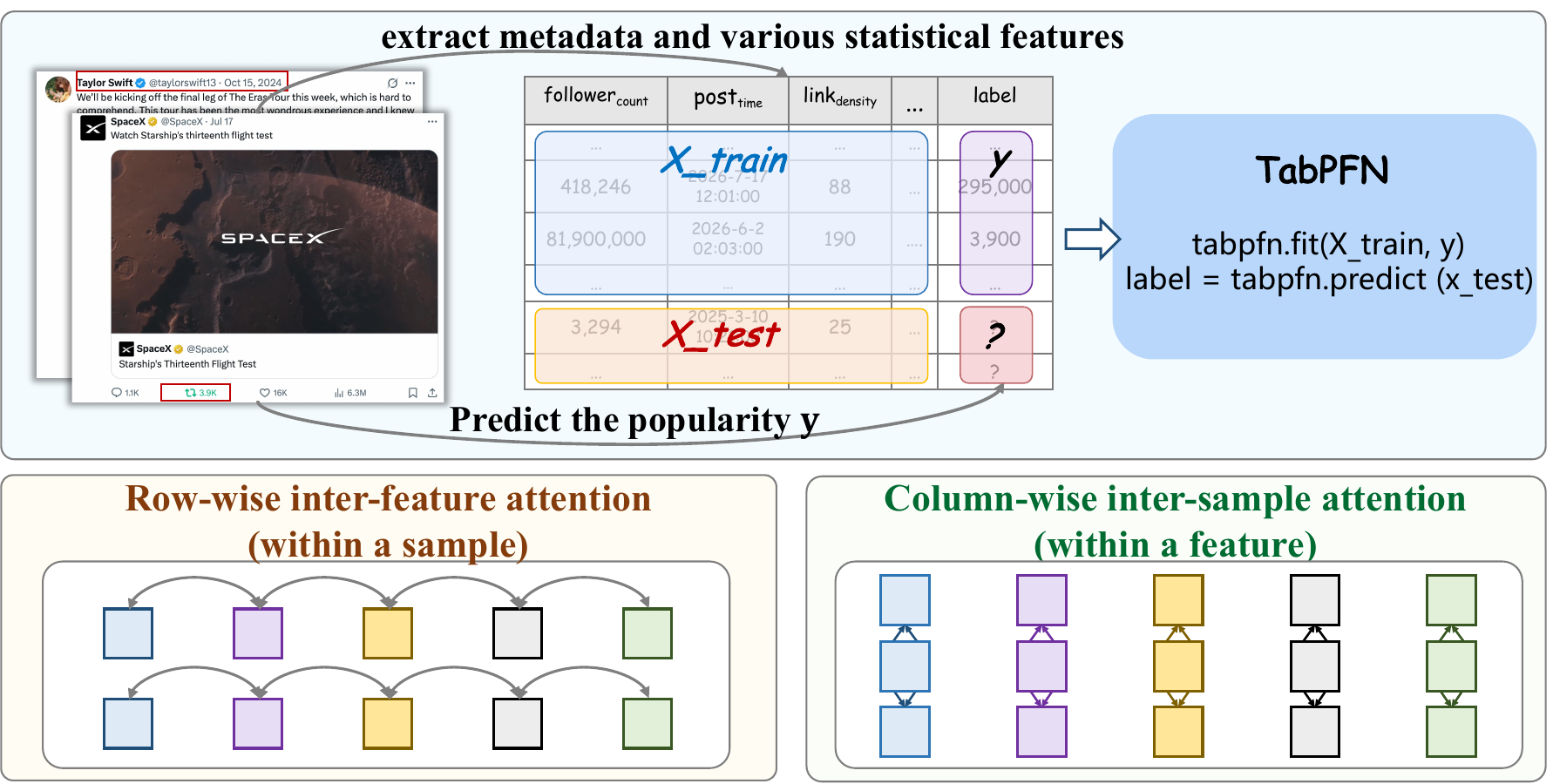}
\vspace{-5mm}
\caption{A toy example of TabPFN for popularity prediction.}
\label{fig:tabpfn}
\vspace{-6mm}
\end{figure}

\subsection{Tabular Foundation Models}
\label{sec:pre_tfm}
Although the formulation above encompasses all four information views, existing methods use them only partially or fuse them shallowly, as discussed in Section~\ref{sec:intro}. Tabular foundation models (TFMs)~\cite{hollmann2022tabpfn, hollmann2025accurate, zhang2025limix} offer a natural substrate for jointly modeling such heterogeneous signals within a single table.


A tabular dataset can be denoted as $\mathcal{D}=(\mathbf{X},\mathbf{y})$, where 
$\mathbf{X}$ comprises $n$ samples (rows) and $d$ heterogeneous features (columns), 
and $\mathbf{y}$ contains the corresponding labels. Throughout this paper, we instantiate TFMs with TabPFN~\cite{hollmann2022tabpfn, hollmann2025accurate, grinsztajn2026tabpfn}, the most representative model of this family. TabPFN adopts a transformer encoder tailored to the two-dimensional structure of tables, natively handling Boolean, categorical, and numerical features as well as missing values. Each cell is first mapped to an embedding and then processed by two-way attention (Fig.~\ref{fig:tabpfn}, bottom): row-wise inter-feature attention captures dependencies among features within a sample, while column-wise inter-sample attention models relationships among samples for a given feature.
TabPFN is pre-trained offline to approximate the Bayesian posterior predictive distribution, using millions of synthetic tabular datasets drawn from a predefined prior. The prior is constructed from Bayesian neural networks~\cite{neal2012bayesian,gal2017uncertainty} and structural causal models (SCMs)~\cite{peters2017elements}, which encode diverse feature dependencies and causal mechanisms underlying tabular data. 
Once pretrained, TabPFN performs in-context learning on an unseen dataset in a single forward pass~\cite{muller2021transformers}: as illustrated in Fig.~\ref{fig:tabpfn} (top), the model infers the feature--label mapping from the labeled context $(\mathbf{X}_{\mathrm{train}},\mathbf{y}_{\mathrm{train}})$ and predicts labels for the test samples $\mathbf{X}_{\mathrm{test}}$, without any parameter update.

\begin{figure*}[t]
\centering
\includegraphics[trim=2 0 0 2, clip, width=0.95\linewidth]{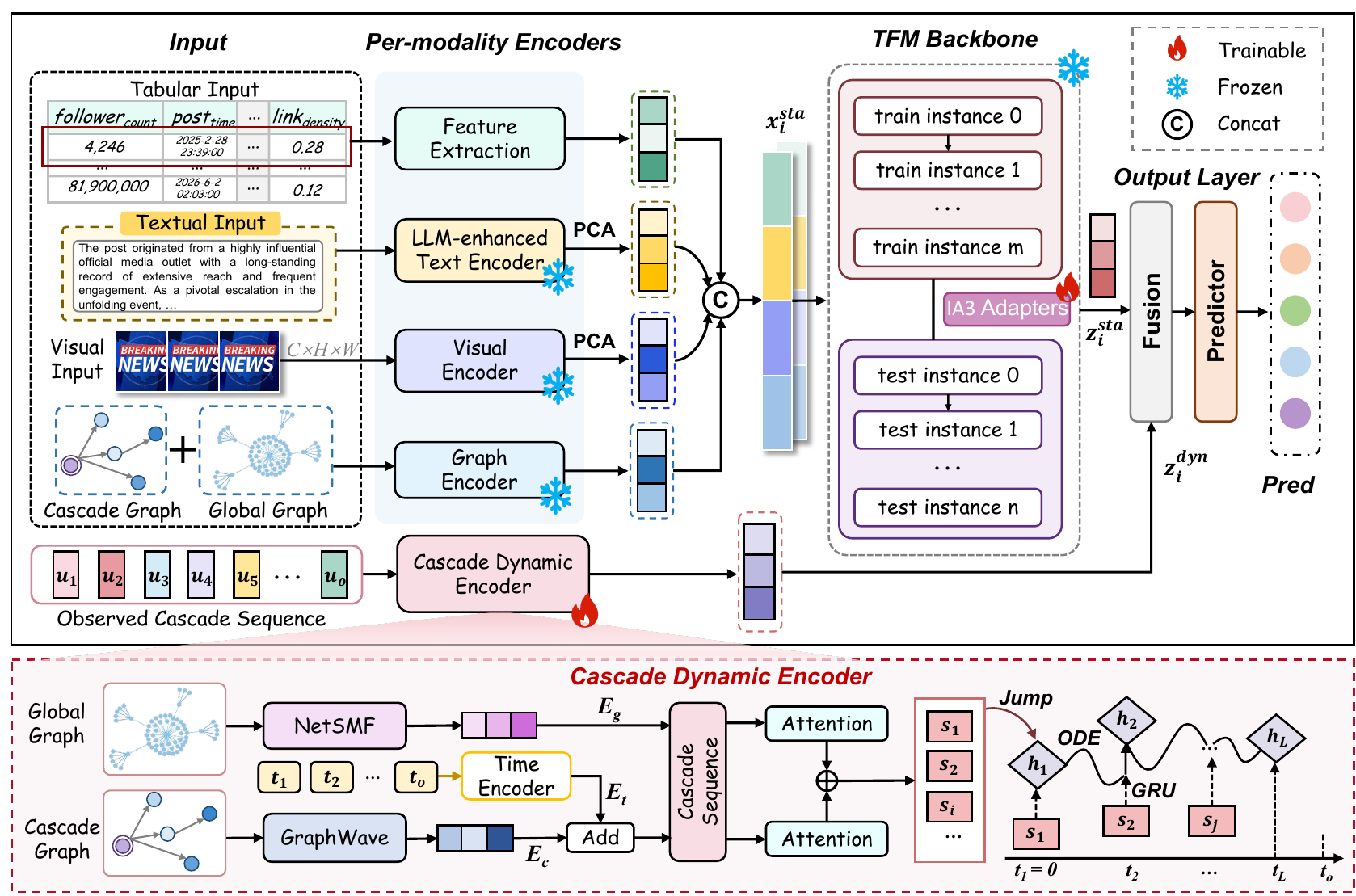}

\vspace{0.3em} 
\caption{\textbf{Overview of TFM4POP.} The static branch assembles the textual, visual, structural, and tabular views into a unified tabular input for the TFM backbone, while the dynamic branch models the continuous-time cascade dynamics with a Neural-ODE-based encoder; the two representations are fused by cross-attention for prediction.}


\label{fig:model_architecture}
\vspace{-2mm}  
\end{figure*}

\section{Methodology}
\label{sec:method}

\subsection{Overview}
\label{sec:overview}

Figure~\ref{fig:model_architecture} illustrates the overall architecture of TFM4POP, a dual-branch design that separates the static attributes of a cascade from its continuous-time diffusion process. In the static branch, frozen per-modality encoders transform the textual, visual, and graph-structural views into compact features, which are concatenated with the raw tabular features into the static tabular input $\mathbf{x}_i^{\mathrm{sta}}$, one row per cascade (Section~\ref{sec:static_tabularization}). The TFM backbone, whose pre-trained weights remain frozen, then performs in-context learning over the resulting table, yielding the static representation $\mathbf{z}_i^{\mathrm{sta}}$; a small set of IA3 vectors adapts it to the cascade domain (Section~\ref{sec:peft}). In the dynamic branch, the cascade dynamic encoder models the observed event sequence with a neural ODE and produces the dynamic representation $\mathbf{z}_i^{\mathrm{dyn}}$ (Section~\ref{sec:dynamics}). The two branches are then fused by a cross-attention module in which the static representation queries the latent trajectory of the dynamic branch, and the fused representation is passed to an MLP predictor that estimates the future incremental popularity $\widehat{\Delta P}_i$ (Section~\ref{sec:fusion}). Only the IA3 vectors, the dynamic encoder, the cross-attention fusion module, and the MLP predictor are trained, using leakage-free out-of-fold context construction (Section~\ref{sec:training}).

\subsection{Static View Encoding with TFMs}
\label{sec:static_tabularization}

The central challenge in applying TFMs to popularity prediction is the modality gap: TFMs are agnostic to unstructured text, images, and graph structures. To bridge this gap, we encode the textual, visual, and structural views into compact feature columns and assemble them with the raw tabular features into a unified table with one row per cascade; the temporal dynamics of the cascade view are handled separately by the dynamic branch (Section~\ref{sec:dynamics}). This allows the TFM to encode all static views uniformly with its pretrained prior, capturing cross-view and cross-sample interactions in a shared tabular space.

\noindent\textbf{Tabular View.}
For each cascade $C_i$, we collect the metadata and statistical features available by the observation time $t_o$ and organize them into a raw tabular feature vector $\mathbf{x}_i^{\mathrm{tab}}\in\mathbb{R}^{d_{\mathrm{tab}}}$. These features characterize four aspects of early cascade diffusion: diffusion volume, user reach and network centrality, propagation depth and breadth, and diffusion speed. All extracted features and their definitions are listed in \textbf{Appendix~\ref{app:tabular_features}}. 


\noindent\textbf{Textual and Visual Views.}
Rather than encoding the original post text $\mathcal{T}_i$ in isolation, we prompt an LLM to assess the post's diffusion potential along four dimensions: poster influence, content virality, posting time, and early-resharer quality, using only information available by $t_o$. The generated rationale consolidates fragmented signals across the poster, the content, and the early diffusion context into a popularity-oriented summary. When the contextual information required for rationale generation is unavailable, we fall back to the original post text. We encode the resulting text with XLM-RoBERTa \cite{conneau2020unsupervised} and mean-pool the valid token representations into the textual feature vector $\mathbf{h}_i^{\mathrm{txt}}$. For a post with visual content $\mathcal{I}_i$, we encode all images or uniformly sampled video frames with a pretrained CLIP image encoder \cite{radford2021learning} and average their embeddings into the visual feature vector $\mathbf{h}_i^{\mathrm{vis}}$; posts without visual content are retained, with their visual features marked as missing.
As both $\mathbf{h}_i^{\mathrm{txt}}$ and $\mathbf{h}_i^{\mathrm{vis}}$ are high-dimensional, naively concatenating them with the raw tabular features would let these views dominate the attention budget and suppress the tabular columns. We therefore apply principal component analysis (PCA) to each view separately, fitting it on the training set and applying it to all splits, yielding compact vectors $\mathbf{x}_i^{\mathrm{txt}}\in\mathbb{R}^{d_{\mathrm{txt}}}$ and $\mathbf{x}_i^{\mathrm{vis}}\in\mathbb{R}^{d_{\mathrm{vis}}}$.

%

\noindent\textbf{Structural View.}
Following prior work~\cite{jing2025casft,jing2026modeling}, we adopt existing graph embedding methods to encode structural information from both the cascade and global graphs. For the cascade graph $G_i(t_o)$, we use GraphWave~\cite{donnat2018learning}, which leverages heat wavelet diffusion patterns to encode the structural role of each participating user, yielding a local embedding $\mathbf{E}_c(u)$ for each user $u$. For the large-scale global graph $\mathcal{G}$, we adopt the fast and scalable NetSMF~\cite{qiu2019netsmf} to obtain a global embedding $\mathbf{E}_g(u)$ that captures the user's interaction patterns across cascades. The resulting user embeddings are also used in the dynamic branch to represent each repost event (Section~\ref{sec:dynamics}). For the static branch, we aggregate the local and global embeddings separately over the users in $G_i(t_o)$, using each user's PageRank score as the corresponding weight. This weighting gives greater importance to structurally influential users. Finally, we concatenate the two aggregated embeddings to form the cascade-level structural feature vector $\mathbf{x}_i^{\mathrm{str}}\in\mathbb{R}^{d_{\mathrm{str}}}$.

\noindent\textbf{Unified In-Context Encoding.}
For each cascade $C_i$, we concatenate the raw tabular features with the extracted textual, visual, and structural features to form its static tabular input:
\begin{equation}
\mathbf{x}_i^{\mathrm{sta}}
=
\mathbf{x}_i^{\mathrm{tab}}
\oplus
\mathbf{x}_i^{\mathrm{txt}}
\oplus
\mathbf{x}_i^{\mathrm{vis}}
\oplus
\mathbf{x}_i^{\mathrm{str}},
\end{equation}
where $\oplus$ denotes feature-wise concatenation. Unlike conventional encoders that process each cascade independently, TabPFN's two-way attention operates over the assembled table: row-wise attention captures cross-view interactions within each cascade, while column-wise attention relates it to similar historical cascades, akin to implicit retrieval augmentation. We take the final-layer hidden state of each query row, i.e., the encoder output immediately before TabPFN's prediction head, as the static cascade representation $\mathbf{z}_i^{\mathrm{sta}}$. As in-context learning conditions on the labeled cascades, we construct the context out-of-fold during training so that a cascade never attends to its own label (Section~\ref{sec:training}).

\subsection{Continuous-Time Dynamics Modeling}
\label{sec:dynamics}

While the static branch summarizes each cascade as a single table row, the diffusion process itself is a continuous-time event sequence in which reposts arrive at irregular intervals, and their timing is a strong indicator of future popularity. Standard RNNs treat events as a uniformly spaced sequence and therefore cannot represent how much time elapses between reposts. We therefore adopt neural ordinary differential equations (ODEs)~\cite{chen2018neural, rubanova2019latent}, which evolve a hidden state in continuous time and naturally accommodate irregularly timed events.

Concretely, for the cascade sequence $C_i(t_o)$ with $L$ observed reposts, we represent each event $(u_j, t_j)$ with the structural embeddings of user $u_j$ obtained in Section~\ref{sec:static_tabularization}: the local embedding $\mathbf{E}_c(u_j)$ from the cascade graph and the global embedding $\mathbf{E}_g(u_j)$ from the global graph. To make the events time-aware, we add a sinusoidal temporal encoding $\mathbf{E}_t$ of the timestamps~\cite{zuo2020transformer} to the local embeddings. Each embedding sequence is then refined by its own causal self-attention module, where each event attends only to its predecessors. The two refined representations are concatenated event-wise to form the mark sequence $\mathbf{S} = (\mathbf{s}_1, \ldots, \mathbf{s}_L)$.
Taking $\mathbf{S}$ as input, we model the cascade dynamics with a hidden state $\mathbf{h}_t$ that evolves continuously between events and jumps at each event arrival. Between two consecutive reposts, $\mathbf{h}_t$ follows an ODE whose drift is parameterized by a fully connected network $f_{\theta}$. When the $j$-th repost arrives at $t_j$, a GRU cell $g$ performs an instantaneous jump update conditioned on the mark $\mathbf{s}_j$:
\begin{align}
\frac{d\mathbf{h}_t}{dt} &= f_{\theta}\left(t, \mathbf{h}_t\right), \\
\mathbf{h}'_{t_j} &= \operatorname{ODESolve}\left(f_{\theta}, \mathbf{h}_{t_{j-1}}, (t_{j-1}, t_j)\right), \\
\mathbf{h}_{t_j} &= g\left(\mathbf{h}'_{t_j}, \mathbf{s}_j\right).
\end{align}
The continuous flow captures the smooth evolution of the cascade over irregular inter-event gaps, while the jumps inject the abrupt changes brought by newly observed reposts. After the last observed event, the state is further propagated by the same ODE to the observation time $t_o$, so that the elapsed silence since the final repost is also encoded. Rather than summarizing the diffusion by its final state alone, we take the full latent trajectory as the dynamic cascade representation,
\begin{equation}
\mathbf{z}_i^{\mathrm{dyn}} = \left(\mathbf{h}_{t_1}, \dots, \mathbf{h}_{t_L}, \mathbf{h}_{t_o}\right),
\end{equation}
which preserves the stage-wise evolution of the cascade and is attended over by the fusion module (Section~\ref{sec:fusion}).

\begin{table*}[t]
\centering
\caption{A brief comparison of the information provided by different cascade popularity prediction datasets. \yesmark\ indicates that the information is provided, whereas \nomark\ indicates that it is unavailable.}
\label{tab:dataset_comparison}
\small
\setlength{\tabcolsep}{4.5pt}
\renewcommand{\arraystretch}{0.85}
\begin{tabular}{@{}lcccccccrrr@{}}
\toprule
Dataset
& Event
& Text
& Image
& Comments
& Social Graph
& User Profile
& User History
& \#Cascades
& \#Users
& Avg.\ Popularity \\
\midrule
Twitter~\cite{weng2013virality}
& \nomark & \nomark & \nomark & \nomark
& \nomark & \nomark & \nomark
& 88,440 & 490,474 & 141.6 \\

Weibo~\cite{cao2017deephawkes}
& \nomark & \nomark & \nomark & \nomark
& \nomark & \nomark & \nomark
& 119,313 & 6,738,040 & 174.0 \\

Twitter (MMCas)~\cite{jing2026modeling}
& \nomark & \yesmark & \yesmark & \nomark
& \nomark & \yesmark & \nomark
& 7,205 & 225,998 & 31.9 \\

\midrule
\textbf{EventCas (Ours)}
& \yesmark & \yesmark & \yesmark & \yesmark
& \yesmark & \yesmark & \yesmark
& 33,455 & 1,109,542 & 270.6 \\
\bottomrule
\end{tabular}
\end{table*}

\subsection{Static-Dynamic Fusion and Prediction}
\label{sec:fusion}
The two branches characterize a cascade from complementary perspectives: $\mathbf{z}_i^{\mathrm{sta}}$ encodes its static multi-view profile, contextualized against similar historical cascades through in-context learning, whereas $\mathbf{z}_i^{\mathrm{dyn}}$ traces the continuous-time evolution of the observed diffusion. Rather than collapsing $\mathbf{z}_i^{\mathrm{dyn}}$ into a single vector for concatenation, we let the static profile actively query the trajectory through cross-attention, attending to the stages of the evolution most relevant to that profile.

Since the two branches produce representations of different dimensionality, we first project the static representation and each state in $\mathbf{z}_i^{\mathrm{dyn}}$ into a shared $d$-dimensional fusion space,
\begin{equation}
\mathbf{q}_i = \mathbf{z}_i^{\mathrm{sta}}\mathbf{W}_{\mathrm{s}},
\qquad
\mathbf{M}_i = \big[\,\mathbf{h}_{t_1}\mathbf{W}_{\mathrm{d}};\ \dots;\ \mathbf{h}_{t_L}\mathbf{W}_{\mathrm{d}};\ \mathbf{h}_{t_o}\mathbf{W}_{\mathrm{d}}\,\big] \in \mathbb{R}^{(L+1)\times d},
\label{eq:fusion_proj}
\end{equation}
and then take the static token as the query and the trajectory tokens as keys and values in multi-head cross-attention, whose per-head projections $\mathbf{W}^{Q},\mathbf{W}^{K},\mathbf{W}^{V}$ operate on this shared space (written in single-head form for clarity):
\begin{equation}
\mathbf{a}_i = \operatorname{MHA}\!\big(\mathbf{q}_i,\ \mathbf{M}_i,\ \mathbf{M}_i\big)
= \operatorname{softmax}\!\Big(\tfrac{(\mathbf{q}_i\mathbf{W}^{Q})(\mathbf{M}_i\mathbf{W}^{K})^{\top}}{\sqrt{d}}\Big)\,\mathbf{M}_i\mathbf{W}^{V},
\label{eq:fusion_xattn}
\end{equation}
The attended vector $\mathbf{a}_i$ thus summarizes the diffusion dynamics conditioned on the static profile. We concatenate it with the projected static token and regress the future incremental popularity with a multilayer perceptron (MLP):
\begin{equation}
\mathbf{z}_i^{\mathrm{fus}} = \mathbf{q}_i \oplus \mathbf{a}_i,
\qquad
\widehat{\Delta P}_i = \operatorname{MLP}\big(\mathbf{z}_i^{\mathrm{fus}}\big).
\label{eq:prediction}
\end{equation}

\subsection{Parameter-Efficient Adaptation}
\label{sec:peft}
The TabPFN backbone is pre-trained purely on synthetic tabular data, whose prior inevitably deviates from the distribution of real-world cascade features; some degree of adaptation is therefore necessary. Fully fine-tuning the backbone, however, would risk catastrophic forgetting of the pre-trained prior on our moderately sized cascade datasets. We instead adopt IA3~\cite{liu2022few}, which keeps all pre-trained weights frozen and inserts learned rescaling vectors: for a frozen linear layer with weight $\mathbf{W}$, the output is modulated channel-wise as
\begin{equation}
\mathbf{y} = \boldsymbol{\ell} \odot \left(\mathbf{x}\mathbf{W}\right),
\label{eq:ia3}
\end{equation}
where $\boldsymbol{\ell}$ is initialized to all ones so that training starts exactly from the pre-trained model. We attach such vectors only to the value projections and feed-forward layers of the last four transformer blocks of the backbone, adapting the most task-specific layers while leaving the early blocks, which encode generic tabular inductive biases, untouched. This updates only about $10^{-4}$ of the backbone parameters, closing the gap 
between the synthetic prior and the cascade domain at negligible cost.

\subsection{Training Objective}
\label{sec:training}
\noindent\textbf{Leakage-Free Out-of-Fold In-Context Training.}
Training the in-context branch requires care: TabPFN conditions on labeled cascades, so naively using the training set as its own context would let each cascade attend to its own label. We therefore partition the training rows into $K$ folds. At each training step, a query batch is drawn from one fold, while the context is assembled exclusively from the remaining $K{-}1$ folds, ensuring that no query ever co-occurs with its own label. At inference, this restriction is lifted and the full training table serves as the context.

\noindent\textbf{Loss.}
Following previous studies~\cite{cao2017deephawkes, xu2021casflow}, we train on log-scaled popularity to counter its heavy-tailed distribution, minimizing
\begin{equation}
\mathcal{L} = \frac{1}{|\mathcal{B}|} \sum_{i \in \mathcal{B}} \left( \log_2\bigl(\widehat{\Delta P}_i + 1\bigr) - \log_2\bigl(\Delta P_i + 1\bigr) \right)^2,
\label{eq:loss}
\end{equation}
where $\mathcal{B}$ denotes a query batch. Under this objective, the IA3 vectors, the Neural-ODE-based dynamic encoder, the cross-attention fusion module, and the MLP predictor are optimized jointly, while the pre-trained modality encoders and TFM backbone remain frozen.

\begin{table*}[t]
    \centering
    \caption{Performance comparisons on Twitter and EventCas under two observation windows, reported as mean $\pm$ standard deviation. Lower MSLE and MAPE indicate better performance. The best and second-best mean results are shown in \textbf{bold} and \underline{underline}, respectively.}
    \label{tab:main_results}
    \begin{threeparttable}
    \setlength{\tabcolsep}{4pt}
    \resizebox{\linewidth}{!}{
    \begin{tabular}{l|cccc|cccc}
        \toprule
        \multirow{3}{*}{\textbf{Model}}
        & \multicolumn{4}{c|}{\textbf{Twitter}}
        & \multicolumn{4}{c}{\textbf{EventCas}} \\
        \cmidrule(lr){2-5}
        \cmidrule(lr){6-9}
        & \multicolumn{2}{c}{1 Hour} & \multicolumn{2}{c|}{3 Hours}
        & \multicolumn{2}{c}{0.5 Hour} & \multicolumn{2}{c}{1 Hour} \\
        \cmidrule(lr){2-3}
        \cmidrule(lr){4-5}
        \cmidrule(lr){6-7}
        \cmidrule(lr){8-9}
        & MSLE~$\downarrow$ & MAPE~$\downarrow$ & MSLE~$\downarrow$ & MAPE~$\downarrow$
        & MSLE~$\downarrow$ & MAPE~$\downarrow$ & MSLE~$\downarrow$ & MAPE~$\downarrow$ \\
        \midrule
        Feature-Linear & 1.6385{\scriptsize$\pm$0.0004} & 0.2523{\scriptsize$\pm$0.0002} & 2.0192{\scriptsize$\pm$0.0005} & 0.3173{\scriptsize$\pm$0.0003} & 5.3156{\scriptsize$\pm$0.0535} & 0.4298{\scriptsize$\pm$0.0029} & 4.2465{\scriptsize$\pm$0.0132} & 0.4028{\scriptsize$\pm$0.0008} \\
        Feature-MLP    & 1.4543{\scriptsize$\pm$0.0149} & 0.2380{\scriptsize$\pm$0.0024} & 1.5845{\scriptsize$\pm$0.0129} & 0.2736{\scriptsize$\pm$0.0013} & 3.9044{\scriptsize$\pm$0.0541} & 0.3512{\scriptsize$\pm$0.0043} & 3.0373{\scriptsize$\pm$0.0611} & 0.3159{\scriptsize$\pm$0.0065} \\
        DeepHawkes     & 1.6442{\scriptsize$\pm$0.0051} & 0.2519{\scriptsize$\pm$0.0045} & 1.9308{\scriptsize$\pm$0.0027} & 0.3070{\scriptsize$\pm$0.0100} & 4.6438{\scriptsize$\pm$0.1648} & 0.3984{\scriptsize$\pm$0.0167} & 3.5093{\scriptsize$\pm$0.3747} & 0.3754{\scriptsize$\pm$0.0342} \\
        UHAN           & 2.1717{\scriptsize$\pm$0.1890} & 0.2779{\scriptsize$\pm$0.0053} & 2.5625{\scriptsize$\pm$0.1108} & 0.3463{\scriptsize$\pm$0.0064} & 5.9482{\scriptsize$\pm$0.0940} & 0.4714{\scriptsize$\pm$0.0130} & 5.6736{\scriptsize$\pm$0.1811} & 0.4798{\scriptsize$\pm$0.0194} \\
        CBAN           & 1.6680{\scriptsize$\pm$0.0212} & 0.2570{\scriptsize$\pm$0.0077} & 2.1104{\scriptsize$\pm$0.0345} & 0.3280{\scriptsize$\pm$0.0090} & 6.3242{\scriptsize$\pm$0.2785} & 0.5706{\scriptsize$\pm$0.0542} & 6.0700{\scriptsize$\pm$0.1571} & 0.5657{\scriptsize$\pm$0.0380} \\
        CasCN          & 1.4677{\scriptsize$\pm$0.0226} & 0.2458{\scriptsize$\pm$0.0046} & 1.7576{\scriptsize$\pm$0.0421} & 0.2981{\scriptsize$\pm$0.0068} & 5.2083{\scriptsize$\pm$0.1777} & 0.4714{\scriptsize$\pm$0.0142} & 4.5882{\scriptsize$\pm$0.1698} & 0.4423{\scriptsize$\pm$0.0259} \\
        CasFlow        & 1.6773{\scriptsize$\pm$0.0302} & 0.2355{\scriptsize$\pm$0.0113} & 1.8310{\scriptsize$\pm$0.0643} & 0.2601{\scriptsize$\pm$0.0102} & 3.7833{\scriptsize$\pm$0.2689} & 0.3014{\scriptsize$\pm$0.0243} & 3.0189{\scriptsize$\pm$0.2386} & \underline{0.2709}{\scriptsize$\pm$0.0101} \\
        MINDS          & 1.7127{\scriptsize$\pm$0.0094} & 0.2586{\scriptsize$\pm$0.0034} & 2.0929{\scriptsize$\pm$0.0297} & 0.3187{\scriptsize$\pm$0.0037} & 5.1774{\scriptsize$\pm$0.3315} & 0.4357{\scriptsize$\pm$0.0291} & 4.0923{\scriptsize$\pm$0.2882} & 0.4267{\scriptsize$\pm$0.0438} \\
        CTCP           & 1.5962{\scriptsize$\pm$0.0156} & 0.2310{\scriptsize$\pm$0.0041} & 1.6604{\scriptsize$\pm$0.0166} & 0.2621{\scriptsize$\pm$0.0057} & 3.9524{\scriptsize$\pm$0.1705} & 0.3380{\scriptsize$\pm$0.0063} & 3.4269{\scriptsize$\pm$0.0741} & 0.3379{\scriptsize$\pm$0.0129} \\
        CasDO          & 1.3619{\scriptsize$\pm$0.0111} & 0.2321{\scriptsize$\pm$0.0013} & 1.4331{\scriptsize$\pm$0.0161} & 0.2548{\scriptsize$\pm$0.0029} & \underline{3.0119}{\scriptsize$\pm$0.0332} & 0.2913{\scriptsize$\pm$0.0088} & 2.5828{\scriptsize$\pm$0.0343} & 0.2847{\scriptsize$\pm$0.0066} \\
        CasFT          & 1.3581{\scriptsize$\pm$0.0070} & \underline{0.2296}{\scriptsize$\pm$0.0010} & 1.3967{\scriptsize$\pm$0.0101} & \underline{0.2532}{\scriptsize$\pm$0.0031} & 3.0847{\scriptsize$\pm$0.1472} & 0.2841{\scriptsize$\pm$0.0116} & 2.5818{\scriptsize$\pm$0.0526} & 0.2858{\scriptsize$\pm$0.0065} \\
        MMCas        & \underline{1.2636}{\scriptsize$\pm$0.0155} & 0.2439{\scriptsize$\pm$0.0032} & \underline{1.3379}{\scriptsize$\pm$0.0189} & 0.2623{\scriptsize$\pm$0.0048} & 3.0142{\scriptsize$\pm$0.1066} & \underline{0.2790}{\scriptsize$\pm$0.0749} &
         \underline{2.5535}{\scriptsize$\pm$0.0912} 
         & 0.2747{\scriptsize$\pm$0.0124} \\
        AutoCas        & 1.4563{\scriptsize$\pm$0.0212} & 0.2402{\scriptsize$\pm$0.0018} & 1.4625{\scriptsize$\pm$0.0311} & 0.2658{\scriptsize$\pm$0.0050} & 3.4493{\scriptsize$\pm$0.1500} & 0.3051{\scriptsize$\pm$0.0086} & 2.7416{\scriptsize$\pm$0.0636} & 0.2948{\scriptsize$\pm$0.0160} \\
        \midrule
        TFM4POP w/ LimiX-16M$^\dagger$ & 1.2839{\scriptsize$\pm$0.0688} & 0.2204{\scriptsize$\pm$0.0265} & 1.2945{\scriptsize$\pm$0.0484} & 0.2387{\scriptsize$\pm$0.0093} & 2.8998{\scriptsize$\pm$0.1537} & 0.2782{\scriptsize$\pm$0.0101} & 2.4224{\scriptsize$\pm$0.1065} & 0.2575{\scriptsize$\pm$0.0106} \\
        \midrule
        \rowcolor{gray!15}
        \textbf{TFM4POP (ours)} & \textbf{1.2382}{\scriptsize$\pm$0.0344} & \textbf{0.2145}{\scriptsize$\pm$0.0029} & \textbf{1.2917}{\scriptsize$\pm$0.0199} & \textbf{0.2459}{\scriptsize$\pm$0.0043} & \textbf{2.7544}{\scriptsize$\pm$0.0932} & \textbf{0.2326}{\scriptsize$\pm$0.0038} & \textbf{2.2276}{\scriptsize$\pm$0.0050} & \textbf{0.2218}{\scriptsize$\pm$0.0021} \\
        \bottomrule
    \end{tabular}}
    \begin{tablenotes}
        \footnotesize
        \item[$\dagger$] This variant replaces the default TabPFN backbone with LimiX-16M while keeping the remaining architecture and training protocol unchanged.
    \end{tablenotes}
    \end{threeparttable}
    \vspace{-2mm}
\end{table*}

\section{Experiments}
\subsection{Evaluation Datasets}
\label{sec:datasets}
We conduct experiments on two real-world datasets: \textbf{Twitter} and \textbf{EventCas}.
The Twitter dataset is introduced by the recent work MMCas~\cite{jing2026modeling} and comprises multimodal cascades with user profiles and textual and visual content, but without comment threads or user histories.
EventCas is a new event-centric multimodal cascade benchmark we curate from Sina Weibo\footnote{\url{https://weibo.com/}}.
Unlike widely used benchmarks such as Weibo~\cite{cao2017deephawkes} and Twitter~\cite{weng2013virality}, which provide only basic cascade data (user IDs and timestamps) without any content or contextual information, EventCas is designed to support multi-view popularity prediction.
Table~\ref{tab:dataset_comparison} compares EventCas with existing datasets, highlighting two key advantages:
\textbf{(1)} it uniquely collects posts associated with specific trending events rather than arbitrarily sampled public posts, grounding each cascade in event context and enabling explicit modeling of event-driven diffusion;
\textbf{(2)} it offers the most comprehensive information coverage among existing datasets, providing the textual and visual content of posts, comment threads, the underlying social graph, user profiles, and user behavior histories, while remaining substantial in scale with 33{,}455 cascades and over 1.1 million users.
The detailed collection process, anonymization, and additional statistics are provided in \textbf{Appendix~\ref{sec:dataset_details}}.

\subsection{Experimental Settings}

\noindent \textbf{Baselines.}
For a comprehensive comparison, we adopt thirteen baselines from four categories.
\textbf{(i)} Feature-based methods: Feature-Linear and Feature-MLP.
\textbf{(ii)} Statistics-based methods: DeepHawkes \cite{cao2017deephawkes}.
\textbf{(iii)} Deep learning-based methods, further divided into content-based approaches, UHAN~\cite{zhang2018user} and CBAN~\cite{cheung2022crossmodal}, and cascade-based approaches, CasCN~\cite{chen2019information}, CasFlow~\cite{xu2021casflow}, MINDS~\cite{jiao2024enhancing}, CTCP~\cite{lu2023continuous}, CasDO~\cite{cheng2024information}, MMCas~\cite{jing2026modeling}, and CasFT~\cite{jing2025casft}.
\textbf{(iv)} LLM-based methods: AutoCas~\cite{zheng2025autocas}.
More baseline details are provided in \textbf{Appendix~\ref{sec:baseline_details}}.

\noindent \textbf{Evaluation Metrics.}
Following prior work \cite{xu2021casflow, cheng2024information, shang2025dvcae}, we use Mean Squared Logarithmic Error (MSLE) and Mean Absolute Percentage Error (MAPE) to evaluate prediction performance.

\noindent \textbf{Implementation Details.} 
All experiments are conducted on a workstation equipped with NVIDIA RTX 5880 Ada GPUs.
Following~\cite{jing2026modeling}, we set the observation time $t_o \in \{1, 3\}$ hours with the prediction time $t_p = 3$ days on Twitter, and $t_o \in \{0.5, 1\}$ hours with $t_p = 24$ hours on EventCas.
We employ Qwen3-Max for LLM-enhanced rationale generation.
For the static views, we set $d_{\mathrm{txt}}=8/4$ and
$d_{\mathrm{vis}}=24/16$ for the PCA-compressed textual and visual features, $d_{\mathrm{str}}=40$ for the aggregated structural features on both datasets, and $d_{\mathrm{tab}}=21/30$ for the raw tabular features.
The hidden dimension of the dynamic encoder is $96/64$, with dropout $0.05/0.1$.
The trainable components are optimized jointly with Adam, using a batch size of 64 and $K=5$ for out-of-fold context construction; the learning rates are $3\times10^{-5}/1\times10^{-4}$ for the IA3 vectors and $3\times10^{-4}/1.5\times10^{-3}$ for the dynamic encoder and prediction head.
All experiments are repeated five times with different random seeds, and we report the mean and standard deviation.

\subsection{Main Results}\label{sec:main_results}
The experimental results comparing \textsc{TFM4POP} with all baselines on Twitter and EventCas are shown in Table~\ref{tab:main_results}. \textsc{TFM4POP} consistently outperforms all baselines across both datasets under all observation settings and metrics, achieving 2.01\%--12.76\% relative improvement on MSLE and 2.88\%--18.12\% on MAPE over the best-performing baselines, with consistent gains across all five random seeds. The improvements are most pronounced on EventCas, whose event-centric collection yields denser textual and visual coverage that the unified tabular space can fully exploit. Moreover, the LimiX-16M variant also surpasses all baselines in most settings, indicating that the framework is agnostic to the TFM backbone and readily benefits from future advances in tabular foundation models.

\subsection{Ablation Study}\label{sec:ablation}

\begin{table}[t]
\centering
\caption{Ablation study on Twitter (1 hour) and EventCas (0.5 hour). 
All results are obtained under 
a single fixed random seed for controlled comparison.}
\label{tab:ablation}
\setlength{\tabcolsep}{3pt}
\resizebox{0.95\columnwidth}{!}{%
\begin{tabular}{lcccc}
\toprule
\multirow{2}{*}{\textbf{Model}}& \multicolumn{2}{c}{\textbf{Twitter}}
  & \multicolumn{2}{c}{\textbf{EventCas}} \\
\cmidrule(lr){2-3}\cmidrule(lr){4-5}
  & MSLE$\downarrow$ & MAPE$\downarrow$
  & MSLE$\downarrow$ & MAPE$\downarrow$ \\
\midrule
\textbf{TFM4POP (Full)}  & \textbf{1.2327} & \textbf{0.2134} & \textbf{2.6613} & \textbf{0.2363} \\
\midrule
\multicolumn{5}{l}{\textit{View contributions}} \\
\midrule
w/o Tabular              & 1.3847 & 0.2364 & 2.9994 & 0.2807 \\
w/o Rationale            & - & - & 2.7247 & 0.2612 \\
w/o Vision               & 1.2349 & 0.2193 & 2.6793 & 0.2553 \\
w/o Structure            & 1.2352 & 0.2173 & 3.1664 & 0.2443 \\
w/o Dynamics             & 1.2425 & 0.2185 & 3.3985 & 0.2517 \\
\midrule
\multicolumn{5}{l}{\textit{Architecture choices}} \\
\midrule
LoRA                     & 1.2373 & 0.2172 & 2.9706 & 0.2398 \\
Full fine-tuning         & 1.2771 & 0.2213 & 2.7404 & 0.2557 \\
w/o IA3 (frozen)         & 1.2354 & 0.2172 & 2.6828 & 0.2804 \\
GRU-only dynamics        & 1.2343 & 0.2174 & 2.6726 & 0.2538 \\
w/o TFM              & 1.4118 & 0.2389 & 4.5445 & 0.4569 \\
Concatenation fusion     & 1.2533 & 0.2196 & 2.6710 & 0.2395 \\
\bottomrule
\end{tabular}%
}
\vspace{-5mm}
\end{table}

To examine the contribution of each component in \textsc{TFM4POP}, we conduct ablation studies on Twitter and EventCas. As reported in Table~\ref{tab:ablation}, we compare the full model with two groups of variants: 
\begin{itemize}[leftmargin=*] 
\item \textbf{View contributions.} \textbf{w/o Tabular}, \textbf{w/o Vision}, and \textbf{w/o Structure} remove the raw tabular attributes, the visual features, and the aggregated structural features from the static tabular input, respectively; \textbf{w/o Rationale} replaces the LLM-generated rationale with the raw post text; \textbf{w/o Dynamics} removes the entire dynamic branch and predicts from the static representation alone. 
\item \textbf{Architecture choices.} For TFM adaptation, \textbf{w/o IA3 (frozen)} keeps the backbone entirely frozen, while \textbf{Full fine-tuning} and \textbf{LoRA} replace IA3 with full backbone updates and with LoRA adapters of matched trainable parameters, respectively. For the remaining designs, \textbf{w/o TFM} replaces the TabPFN backbone with a parameter-matched MLP over the same static input; \textbf{GRU-only dynamics} degrades the inter-event ODE flow to an identity transition while keeping the GRU jump updates; and \textbf{Concatenation fusion} replaces the cross-attention fusion with direct concatenation of the two representations. \end{itemize}
 
First, removing any single information view consistently degrades performance, with MSLE increasing by up to 27.70\%, confirming that the four views provide complementary signals. Second, replacing the TFM backbone with a parameter-matched MLP causes the largest degradation among all variants (14.53\% and 70.76\% MSLE increases on Twitter and EventCas), demonstrating that the improvement stems from the pre-trained tabular prior rather than merely from the assembled multi-view features. Third, IA3 outperforms full fine-tuning, LoRA, and the frozen backbone while updating only about $10^{-4}$ of the backbone parameters, corroborating that full-parameter updates risk catastrophically forgetting the pre-trained prior. Finally, degrading the ODE flow to GRU-only updates or replacing cross-attention with concatenation both hurt performance, validating the continuous-time dynamics modeling and the fusion design.


\subsection{Hyperparameter Sensitivity Analysis}
\begin{figure}[t]
\centering
\includegraphics[
trim=4 4 5 3, clip,
  width=\columnwidth,
  height=0.3\textheight,
]{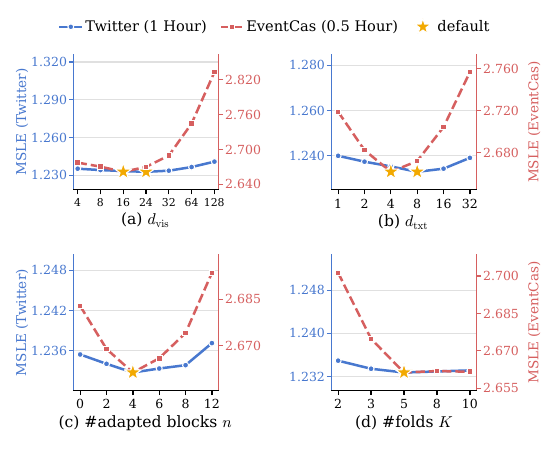}
\vspace{-5mm}
\caption{Hyperparameter sensitivity of \textsc{TFM4POP} w.r.t. (a) the
visual PCA dimension $d_{\mathrm{vis}}$, (b) the textual PCA dimension
$d_{\mathrm{txt}}$, (c) the number of IA3-adapted ICL blocks $n$
($n{=}0$ denotes the frozen backbone), and (d) the fold number $K$,
varying one hyperparameter at a time. Stars mark each dataset's
default configuration.}
\label{fig:param_sensitivity}
\vspace{-5mm}
\end{figure}

We study the sensitivity of \textsc{TFM4POP} to four key hyperparameters, varying one at a time while fixing the others at each dataset's default configuration (Figure~\ref{fig:param_sensitivity}). \textbf{(a--b) PCA dimensions.} Performance follows a U-shaped trend for both $d_{\mathrm{vis}}$ and $d_{\mathrm{txt}}$: overly aggressive compression discards informative signals, while overly large dimensions let the visual and textual columns crowd out the tabular attributes, degrading accuracy and supporting the attention-imbalance concern raised in Challenge~1 (\textbf{C1}). The optima differ across datasets ($d_{\mathrm{vis}}{=}24/16$ and $d_{\mathrm{txt}}{=}8/4$ on Twitter/EventCas), and EventCas is visibly more sensitive than Twitter, echoing the view-contribution results in Section~\ref{sec:ablation}. \textbf{(c) IA3 depth.} Adapting the last four ICL blocks yields the best results on both datasets; the frozen backbone ($n{=}0$) underperforms, while adapting all blocks brings no further gain and slightly erodes the pre-trained prior. \textbf{(d) Fold number.} Performance degrades noticeably only when $K\leq 3$ shrinks the available in-context examples, and remains stable for $K\geq 5$, indicating that the out-of-fold context construction requires no careful tuning.



\section{Conclusion}
In this paper, we present \textsc{TFM4POP}, to the best of our knowledge, the first framework that introduces tabular foundation models into information cascade popularity prediction. By compressing the textual, visual, and structural views into compact feature columns and assembling them with the raw tabular attributes into a unified tabular space, \textsc{TFM4POP} enables a pre-trained TFM to jointly reason over all static views via in-context learning. A dedicated Neural ODE-based dynamic branch further captures the continuous-time cascade dynamics, and a cross-attention module fuses the two complementary representations. To adapt the TFM to real cascade distributions without forgetting its pre-trained prior, we apply parameter-efficient IA3 fine-tuning, which matches or exceeds full fine-tuning while updating only about $10^{-4}$ of the backbone parameters. In addition, we construct and release EventCas, an event-centric multi-view cascade benchmark with the most comprehensive information coverage among existing cascade benchmarks. Extensive experiments on two real-world datasets show that \textsc{TFM4POP} consistently outperforms state-of-the-art baselines, achieving up to 12.76\% and 18.12\% improvement on MSLE and MAPE, respectively. Ablation, sensitivity, and data-efficiency analyses further validate each design choice and the value of the pre-trained tabular prior.

\bibliographystyle{ACM-Reference-Format}
\bibliography{references}

\appendix

\section{Dataset Details}
\label{sec:dataset_details}

\subsection{Limitations of Existing Datasets.}
Widely used popularity prediction benchmarks such as Weibo~\cite{cao2017deephawkes} and Twitter~\cite{weng2013virality} only contain a social graph and basic cascade metadata (user IDs and timestamps), but do not include any textual data that modern LLMs can leverage (e.g., post content, repost text, comments, and user-generated context). Moreover, these datasets are typically constructed by randomly sampling public posts within a fixed time span, which weakens the linkage to concrete real-world events and limits the availability of event-level context.
As a result, models trained on these datasets can hardly capture richer contextual factors that motivate user participation, such as user preferences, content semantics, and evolving event narratives.
To enable evaluation under more realistic, context-rich diffusion settings, we construct \textbf{EventCas}, a new event-centric dataset for popularity prediction that augments cascades with comprehensive contextual information, including post contents, comments, repost texts, user profiles, user behavior histories, and the underlying social graph.

\subsection{Data Construction and Statistics.}
EventCas is collected from Sina Weibo\footnote{\url{https://weibo.com/}}, the most popular microblogging platform in China.
We curate 148 hot events over the past two years, covering diverse topics such as entertainment, sports, social incidents, and public affairs.
For each event, we crawl posts containing event-specific hashtags or keywords, and keep posts with more than five retweets.
For every retained post, we record the user IDs and timestamps of all retweets and construct the retweet cascade by sorting retweets chronologically.
To recover the retweet structure, we parse repost texts using the pattern \texttt{//\@username:} to extract intermediate retweet chains.
For example, if user $B$ reposts a post originally published by user $D$ and the repost text contains \texttt{...//\@C:...//\@A:...}, we infer the repost path as $D \rightarrow A \rightarrow C \rightarrow B$.
In addition, we collect the original post content, comment threads, repost texts, user profiles and users' historical posting behaviors.
To build the underlying social graph, we further crawl follower--followee relationships among all involved users.
Overall, EventCas contains 33,455 posts associated with the selected events, involving 1,109,542 unique users. Each post receives an average of 270.6 reposts, and the dataset contains 2,269,924 comments in total.

\subsection{Dataset Anonymization.}
We anonymize user information to protect privacy while preserving diffusion structure and social connectivity.
Specifically, we replace potentially identifying fields (e.g., user IDs, usernames, and post IDs) with new pseudonymous identifiers via a consistent one-to-one mapping.
For textual content, we also substitute all mentioned usernames with their anonymized counterparts.
These steps conceal personally identifiable information while maintaining the integrity of cascade traces, repost paths, and follower--followee relations through the anonymized identifiers.

\section{Baseline Details}
\label{sec:baseline_details}

\subsection{Feature-based methods} 
\textbf{Feature-P} adopts the early-popularity assumption, where the observed cascade size at the observation time $t_o$ is used as a predictive signal to estimate the popularity at a future time $t_p$, i.e., predicting $P(t_p)$ from $P(t_o)$~\cite{szabo2010predicting}.
In addition, following prior work~\cite{xu2021casflow}, we extract a set of handcrafted structural and temporal features from cascades and feed them into two predictors:
\textbf{Feature-Linear}, which employs a linear regression model, and
\textbf{Feature-MLP}, which replaces the linear predictor with a two-layer multilayer perceptron while using the same input features.

\subsection{Statistical-based methods} 

\noindent \textbf{DeepHawkes}~\cite{cao2017deephawkes} represents each cascade as a set of diffusion paths and employs recurrent neural networks to model the self-exciting dynamics of reposting events, jointly learning temporal decay and user influence factors to predict future cascade growth.

\subsection{Deep learning-based methods} 
\textbf{CasCN} \cite{chen2019information} models a cascade as a time-ordered sequence of sub-cascade graphs, uses graph convolution to encode each snapshot’s local structure, and an LSTM to capture the structural evolution over time for popularity prediction.

\noindent \textbf{CasFlow} \cite{xu2021casflow} models cascade diffusion in a hierarchical manner by jointly capturing local and global structural patterns, and accounts for propagation uncertainty to improve cascade growth prediction.

\noindent \textbf{MINDS} \cite{jiao2024enhancing} jointly models macroscopic and microscopic diffusion by constructing sequential hypergraphs to capture cross-cascade interactions over time, and learns shared representations to enhance multi-scale diffusion prediction.

\noindent \textbf{CTCP} \cite{lu2023continuous} models cascade popularity prediction as a continuous-time graph learning problem and maintains dynamically evolving representations for users and cascades to capture cross-cascade correlations and continuously changing user preferences.

\noindent \textbf{CasDO} \cite{cheng2024information} models information diffusion in continuous time by combining neural ordinary differential equations with probabilistic diffusion models, explicitly capturing temporal irregularity and propagation uncertainty for cascade popularity prediction.

\noindent \textbf{CasFT} \cite{jing2025casft} models the future popularity trend after the observation time by extracting dynamic cues of the growth rate with neural ODEs, and uses them together with the observed cascade representation as conditions to guide a diffusion model that generates future popularity-increasing trends for prediction.

\noindent \textbf{MMCas} \cite{jing2026modeling} models information cascades in a multimodal manner by jointly encoding cascade dynamics, user profiles, textual and visual content, and designs a mixture-of-experts interaction mechanism with a reweighting module to fuse these modalities and provide interpretable popularity prediction.

\noindent \textbf{AutoCas} \cite{zheng2025autocas} models cascade diffusion as an autoregressive sequence generation process by tokenizing cascades into local- and global-aware tokens aligned with the embedding space of large language models, and introduces cascade prompt learning to adapt the LLM backbone to popularity prediction.

\section{Tabular Feature Definitions}
\label{app:tabular_features}

Table~\ref{tab:tabular_features} lists all tabular features extracted from the
early cascade dynamics observed within the observation window~$t_o$, organised
by the four semantic aspects.
All features are computed solely from information available at~$t_o$.

\begin{table*}[t]
\centering
\small
\renewcommand{\arraystretch}{1.15}
\begin{tabular}{p{4.5cm} p{12.5cm}}
\hline
\multicolumn{2}{c}{\textbf{Diffusion Volume}} \\
\hline
$\mathit{obs\_repost\_count}$ & Total number of reshares observed within $[0, t_o]$ \\
\hline
\multicolumn{2}{c}{\textbf{User Reach and Network Centrality}} \\
\hline
$\mathit{poster\_fan\_count}$& Follower count of the original poster \\
$\mathit{poster\_pagerank}$
  & PageRank score of the original poster in the global social graph (log-transformed);reflects the poster's network-wide influence \\
$\mathit{obs\_repost\_user\_avg\_fan\_count}$
  & Average follower count of all users who reshared in $[0, t_o]$ \\
$\mathit{obs\_repost\_user\_avg\_pagerank}$
  & Mean PageRank score (log-transformed) of resharing users in $[0, t_o]$ \\
$\mathit{obs\_repost\_user\_median\_pagerank}$
  & Median PageRank score (log-transformed) of resharing users in $[0, t_o]$ \\
\hline
\multicolumn{2}{c}{\textbf{Propagation Depth and Breadth}} \\
\hline
$\mathit{cascade\_depth}$
  & Longest root-to-resharer path length (in edges) in the cascade tree \\
$\mathit{avg\_branching\_factor}$
  & Average out-degree of non-leaf nodes in the cascade tree \\
$\mathit{depth\_prime\_k}$
  & Slope of the origin-anchored least-squares fit of the per-reshare depth sequence
    against reshare index, $\beta = \sum_i i\,d_i / \sum_i i^2$;
    larger values indicate chain-like propagation \\
$\mathit{depth\_avg\_k}$
  & Mean tree depth across all reshares in $[0, t_o]$ \\
$\mathit{depth\_90p\_k}$
  & 90th-percentile tree depth across all reshares in $[0, t_o]$ \\
\hline
\multicolumn{2}{c}{\textbf{Diffusion Speed}} \\
\hline
$\mathit{hour}$
  & Hour of day (0--23) when the original post was published \\
$\mathit{day}$
  & Day of week when the original post was published (0\,=\,Monday, 6\,=\,Sunday) \\
$\mathit{median\_repost\_interval}$
  & Median inter-event time (seconds) between consecutive reshare events in $[0, t_o]$ \\
$\mathit{time\_prime\_first\_half}$
  & Mean inter-reshare interval in the first half of $[0, t_o]$,
    $\frac{1}{k/2-1}\sum_{i=1}^{k/2-1}(t_{i+1}-t_i)$ \\
$\mathit{time\_prime\_second\_half}$
  & Mean inter-reshare interval in the second half of $[0, t_o]$,
    $\frac{1}{k/2-1}\sum_{i=k/2}^{k-1}(t_{i+1}-t_i)$ \\
$\mathit{time\_double\_prime\_k}$
  & Slope of the origin-anchored least-squares fit of inter-event intervals against
    reshare index, $\beta = \sum_i i\,\Delta t_i / \sum_i i^2$;
    $\beta>0$ indicates deceleration, $\beta<0$ indicates acceleration \\
$\mathit{time\_since\_last\_repost}$
  & Seconds from the last observed reshare to window end $t_o$;
    equals $t_o$ when no reshare occurs, approaches $0$ when the cascade is still
    active near $t_o$ \\
$\mathit{last\_third\_repost\_ratio}$
  & Fraction of reshares in the final third $(2t_o/3,\,t_o]$ relative to all reshares
    in $[0, t_o]$; smaller values indicate earlier activity decay \\
$\mathit{late\_to\_early\_velocity\_ratio}$
  & Ratio of reshare count in $(2t_o/3,\,t_o]$ to count in $(0,\,t_o/3]$;
    values $>1$ indicate late-stage acceleration, $<1$ indicate cooling;
    capped at $10$ when the early window contains no reshares \\
$\mathit{inter\_event\_acceleration}$
  & Mean second-order difference of inter-event intervals,
    $\frac{1}{k-2}\sum_{i=1}^{k-2}(\Delta t_{i+1}-\Delta t_i)$;
    positive indicates growing gaps (deceleration), negative indicates acceleration;
    set to $0$ when $k < 4$ \\
\hline
\end{tabular}
\caption{Tabular features $\mathbf{x}_i^{\mathrm{tab}}$ extracted from early cascade
  dynamics within observation window~$t_o$, organised by the four semantic aspects. All values are computed solely from
  information available at~$t_o$. The table lists the 21 features shared by both
  the Weibo and Twitter datasets; Weibo additionally includes poster/resharer
  historical activity features and social-neighbourhood features unavailable in the
  Twitter data.}
\label{tab:tabular_features}
\end{table*}

\end{document}